\documentclass[12pt,a4paper,final]{iopart}
\usepackage{graphicx}
\usepackage{dcolumn}  
\usepackage{bm}       
\usepackage{amssymb} 
\usepackage{braket}
\usepackage{mathcomp}

\begin{document}

\title[Preparing an article for IOP journals in  \LaTeXe]{AdS/CFT correspondence: the fountain of quantum youth}
\vspace{3pc}

\author{Ovidiu Racorean}
\address{General Direction of Information Technology}
\address{Banul Antonache str. 52-60, sc.C, ap.19, Bucharest, Romania}
\ead{ovidiu.racorean@mfinante.gov.ro}
\vspace{3pc}

\begin{abstract}
\vspace{1pc}

We argue, in the context of AdS/CFT correspondence, that the structure of the geometry dual to two entangled CFTs is a time non-orientable spacetime. Further, we elevate this argument to whatever entangled quantum systems. Accordingly, we should expect that entangled quantum systems (particles in subsidiary) to not experience the flow of time. As a result, the lifetime of entangled particles should be considerably longer than that of their unentangled counterparts.  

\end{abstract}

%\pacs{04.70.Dy, 03.67.Bg, 42.50.Ex, 95.30.Gv}
\vspace{7pc}
\noindent{}      \hspace{7pc}          Essay written for the Gravity Research Foundation 
\vspace{1pc}

\noindent{}       \hspace{10pc}              2020 Awards for Essays on Gravitation
\vspace{2pc}

\noindent{}  \hspace{15pc}  March 13, 2020
%\submitto{ Essay written for the Gravity Research Foundation 2017 Awards for Essays on Gravitation}
\maketitle

Quantum entanglement is one of the most counter-intuitive properties of quantum systems. Einstein, Podolsky and Rosen used entanglement to argue that quantum mechanics predicts a breakdown in locality \cite{epr}. In recent years, the concept of entanglement is applying into the understanding of phenomena related the to high energy and gravity physics. This leads to important discoveries about the connection of entanglement and geometry. It has been proposed in \cite{mal}, \cite{raa}, that spacetime connectedness in AdS is related to quantum entanglement in the dual field theory. Moreover, this occurrence was elevated in \cite{sus}, \cite{jen}, \cite{son} to a general principle: whenever two quantum systems are entangled, there should be some version of a physical spacetime connection between them.

In this essay we will argue that if this scenario is correct, than the spacetime connecting the two quantum systems should be time non-orientable. This statement is based on another interesting insight into the phenomenology of quantum entanglement that comes from emergence of the thermodynamic arrow of time \cite{par}, \cite{mac}, \cite{mic}, in the presence of initial high-correlations between the quantum systems.  On this line of work, we previously argued \cite{
rac}, in the context of AdS/CFT correspondence, that the global spacetime dual to quantum entanglement in the field theory must be time non-orientable.

The argument of time non-orientability of the global spacetime dual to any entangled quantum systems has one notable consequence: the entangled pair is not experiencing the flow of time. As a result, the lifetime of quantum entangled particles should be considerably longer than their unentangled counterparts. 

To motivate this result, let us start considering two non-interacting copies of CFT on sphere $S^d$ noted left ($L$) and right ($R$), such that we can decompose Hilbert space $H_{LR}$ of the composite system as  $H_{LR}=H_L\otimes H_R$. We assume that the two subsystems are initially uncorrelated to begin with, such that the joint state of the system is the product state, $\rho_{LR}=\rho_L\otimes \rho_R$, with $\rho_L$, $\rho_R$ and $\rho_{LR}$ the density matrix of the left, right ant the composite system, respectively, corresponding to the thermal states of the field theory. From this initial product state the correlations between the two quantum systems can only increase. 

We employ here as a natural measure of the correlations between the two subsystems the mutual information:

\begin{equation}
I(\rho_{LR} )= S(\rho_L )+S(\rho_R )-S(\rho_{LR} ),
\end{equation}

where $S(\rho_L)$, $S(\rho_R)$ and $S(\rho_{LR})$ are the entropies of the left, right and the composite system, respectively, defined as the von Neumann entropy, $S(\rho)=-Tr(\rho log\rho)$.

Since we are in a low-entropy environment,  $I(\rho_{LR} )= 0$, which reduces the Eq. (1) to: 

\begin{equation}
S(\rho_{LR})=S(\rho_L)+S(\rho_R).
\end{equation}

As a result, the mutual information and consequently the entropy of the composite system can only increase. To see this, let us now consider entangling some of the degrees of freedom of the individual components, evolving in this way the entropy of the joint state from the initial,  $S_i (\rho_{LR} ) = S_i (\rho_L )+S_i (\rho_R )$,	 to the final entropy  $S_f (\rho_{LR} ) \leq S_f (\rho_L ) + S_f (\rho_R )$.					
The initial and final states of the composite system, are related unitarily, thus $S_f (\rho_{LR} )=S_i (\rho_{LR} )$, such that we have

\begin{equation}
\Delta S(\rho_L )+\Delta S(\rho_R )\geq 0.
\end{equation}

This result is consistent with the second law of thermodynamics which states that the entropy of an isolated system can only increase. In this case, in accordance with the formulation of the second law, the flow of time is directed toward the standard thermodynamic arrow.

On the gravity side of the AdS/CFT correspondence, the interpretation of this initially uncorrelated state is straightforward \cite{raa}. The two separate physical systems determined by the density matrix $\rho_L$ and $\rho_R$, correspond in the dual geometric description, to a global time-oriented spacetime having two separate asymptotically AdS spacetime. 

We consider now that the two copies of the field theory are high-correlated, to begin with. In this scenario the joint state of the two subsystems can be represented, as $\rho_{LR}=\ket{\Psi_\beta} \bra{\Psi_\beta}$, with $\ket{\Psi_{\beta}}$  defined as the thermofield double state,

\begin{equation}
\ket{\Psi_{\beta}} = \frac{1}{\sqrt{Z_{\beta}}}\sum_ne^\frac{-\beta E_n}{2}{\ket{E_n}}_L{\ket{E_n}}_R,
\end{equation}

where  $\ket{E_n}_L(\ket{E_n}_R)$ is the n-th energy eigenvector for the subsystem $L(R)$ , $\beta$ is the inverse temperature and ${Z_\beta}^{(-1⁄2)}$ is a normalization constant. We may remark here that the two subsystems are initially entangled in a pure state. Note however that the individual states of the two subsystems are the thermal states. 

In stark contrast to initially uncorrelated case, we can emphasize here that the initial pure state of the two copies of the field theory ensures that $\rho_L$ and  $\rho_R$ are isospectral so that $S(\rho_L )=S(\rho_R )$. Let us now start from the thermofield double state and gradually disentangle the degrees of freedom. The initial entanglement of the composite system forces the individual entropies $S(\rho_L)$ and $S(\rho_R )$ to move in the same direction, such that, $\Delta S(\rho_L ) = \Delta S(\rho_R)$, at all times. In addition, we can say that the initial pure state, implies that the individual entropies can only decrease, such that $\Delta S(\rho_L ) = \Delta S(\rho_R )\leq 0$. 

In this case, Eq.(3) is reversed as:

\begin{equation}
\Delta S(\rho_L )+\Delta S(\rho_R )\leq 0,
\end{equation} 

a result which suggests that ,from the perspective of the second law of thermodynamics, both orientations of the thermodynamic arrow are allowed, such that there is no opportunity for the dominance of one direction of time over the other \cite{par}, \cite{mac}, \cite{mic}.  

The initial high correlations of the two CFT’s ensure that in the dual spacetime the arrow of time may be oriented normal, from the past to the future or in reverse, from the future to the past. In this context, since there is no preferred orientation of time on the spacetime, we may conjecture that the gravity dual is a time non-orientable spacetime, \cite{rac}.

Let us now consider a scenario in which we start with the two copies of the field theory entangled in the thermofield double state and gradually disentangle the degrees of freedom to zero. It has been pointed out in \cite{raa}, that as the entanglement between L and R CFTs decreases to zero, the mutual information also goes to zero, $I(\rho_LR )= 0$. As a result, we remain in the end with two completely separate physical systems which do not interact, such that the joint state of the system is the product state  $\rho_{LR}=\rho_L \otimes \rho_R$, as in the case of initially uncorrelated CFTs.

From the gravity dual perspective, the above scenario is translated in the following statement: starting with a global time non-orientable spacetime and decreaseing to zero the entanglement between the dual copies of CFT, the resulting spacetime is time orientable. Roughly speaking, disentangling the degrees of freedom between two copies of field theory implies, on the geometry side, a transition from a time non-orientable spacetime to a spacetime having a definite orientation of time, thus a time-orientable spacetime. 

Let us now assume in the spirit of \cite{sus}, \cite{jen}, \cite{son} that the structure of global spacetime play a role in the non-gravitational systems. Thus even the singlet state of particles pair spins should be related by some version of a physical spacetime connection between them. As we have argued in the discussion above the global spacetime the entangled pair lives in must be a time non-orientable spacetime. This statement as abstract as it might be does have remarkable consequences. 

As a result of the absence of a preferred orientation of time on time non-orientable spacetime manifolds, the entangled pair ensemble does not evolve in time. In other words, the entangled pair is not experiencing the flow of time. We can think of some properties devolving from here, like a possible resolution of the non-locality of entangled quantum states. Here we like to draw the reader attention to another particularity implied by the time non-orientability of the spacetime dual to entangle d quantum systems. The absence of the flow of time is reflected in the lifetime of the entangled pair. Accordingly, when particles are considered, the lifetime of quantum entangled particles should be considerably longer than the lifetime of their unentangled counterparts.

To see this more clearly, let us destroy (by some measurements) the entanglement between the pair of particles. From the point of view of the geometry, there is a transition to a time orientable spacetime, in such a way that the unentangled particles live now on a time-orientable spacetime.   The, now, unentangled pair will experience the flow of time and consequently all phenomenon associated to motion. 

We conclude the essay by pointing out that measuring the lifetime of entangled particle compared to the lifetime of the same unentangled particles could shed light on some theoretical aspects of AdS/CFT duality.

\end{document}